\documentclass[aps,onecolumn,groupedaddress,floatfix]{revtex4}
\usepackage[colorlinks=true,citecolor=red,filecolor=green,linkcolor=blue,pdfnewwindow=true]{hyperref}
\usepackage{amsmath} 
\usepackage[version=4]{mhchem}
\usepackage{color}
\usepackage{graphicx}
\usepackage{makeidx}
\usepackage{bm}
\usepackage[title]{appendix} 
\usepackage{tabu}
\usepackage{float}
\usepackage{caption}
\usepackage{subcaption}
\usepackage{hyperref}
\usepackage[normalem]{ulem}
\usepackage{natbib}
\usepackage{threeparttable}
\usepackage[dvipsnames]{xcolor}

\hypersetup{urlcolor=blue}

\makeatletter
 
\newcommand{\Rmnum}[1]{\expandafter\@slowromancap\romannumeral #1@}
\makeatother
\begin{document}

\title{Spatiotemporal patterns of Covid-19 pandemic in India: Inferences of pandemic dynamics from data analysis.}

\author{Preet Mishra and R.~K.~Brojen Singh}
\email{brojen@jnu.ac.in}
\affiliation{School of Computational \& Integrative Sciences, Jawaharlal Nehru University, New Delhi-110067, India.}

\begin{abstract}
\noindent Modeling and analysis of the large scale Covid-19 pandemic data can yield inferences about it's dynamics and characteristics of disease propagation. These inferences can then be correlated with contextual factors like population density, effects of strategic interventions, heterogeneous disease propagation etc, and such set of validated inferences can serve as precedents for designing of subsequent mitigation strategies. In this work, we present the analysis of Covid-19 pandemic data in Indian context using growth functions fitting procedure and harmonic analysis method. Our results of growth function fitting to the data indicate that the growth function parameters are quite sensitive to the growth of the infected population indicating positive impact of lockdown strategy,  identification of inflection point and nearly synchronous statistical features of disease spreading. The harmonic analysis of the data shows the countrywide synchronous incident features due to simultaneous implementation of control strategies. However, if one analyzes the data from each state of the India, one can see various forms of travelling waves in the countrywide wave pattern. Hence, one needs to do these analysis from time to time to understand the effectiveness of any control strategy and to closely look at the disease propagation to devise the required type of mitigation strategies.\\

{\noindent}{{\it \textbf{Keywords:}} Covid-19; Growth functions; Harmonic analysis; Empirical Mode Decomposition(EMD); Control strategy.}

\end{abstract}

\maketitle


\section{Introduction}
\noindent The ongoing Covid-19 pandemic is a major health threat to the human population across the world \cite{Wang}. The threat is quite dangerous just like arms conflict \cite{Marciano}, and can cause serious problems in various perspectives, such as, unstoppable infections due to genetically driven continuous emergence of multiple variants \cite{Bano}, more dangerous forms of Covid-19 disease when it aligns with co-morbidities namely, malaria \cite{Mohanan},  tuberculosis \cite{Maia},  monkeypox \cite{Leon}, cancer \cite{Yang}, diabetes \cite{Steenblock}, cardiovascular diseases \cite{Amankwaah}, other viral diseases \cite{Wilder} etc to cause major harm in global economic crisis \cite{Morgan}, collapse of healthcare systems \cite{Christopher}, risk due to global environmental change \cite{Barouki}, harm in the tourism policies \cite{LiZ}, major loss in agricultural products and policies \cite{Lioutas}, instabilities in socio-psychological equations \cite{Jaspal} etc. These problems will persist until ongoing pandemic is controlled or the disease is cured. Even though few vaccines are available \cite{Ndwandwe}, it is still hard to control the pandemic due to the emergence of new variants, and the psychological fear still remains due to the possibilities of emergence of new variants in future.\\

{\noindent}The problem that we are still facing is how to optimally control the pandemic or cure the disease. One straightforward way to control this pandemic is to hunt for drug/s to kill the virus, SARS-CoV-2 and its variants \cite{Otto} which is still not successful. To control the pandemic, besides the large scale vaccination \cite{Dinleyici}, there have been various strategies, such as lockdown, social distancing, wearing mask etc at various social gatherings, but the pandemic is still hard to control \cite{Perra}. One of the reasons for inability to control this pandemic could be problems in proper understanding of the dynamics of the pandemic and its causative agent. In this perspective, modeling of the disease spreading scenario data, their patterns and spatiotemporal attributes could probably give more concrete answers. Modeling the nature and growth of numbers of an epidemic/pandemic outbreak is important for optimizing the tangible solutions needed to contain and mitigate the epidemic/pandemic. In India, a multi-step approach to design a comprehensive strategic response to the risk posed by COVID-19 caused by SARS-CoV-2 is in progress and implemented time to time as government policy by looking at the patterns of disease spreading \cite{Rohith2020,Biswas2020,Tiwari2020}. With the current wide-scale biotechnology-aided designing of strategies and solutions (e.g. vaccine, anti-viral drugs), and by systematic analysis of spatiotemporal disease spreading patterns, there is a high probability that the risk will be lowered significantly in near future.\\

\noindent Working under constraints of economic, health-system’s carrying capacities, socio-psychotic fear and drastically affected food products supply, one of the major aims of a strategic optimal \cite{Zadeh1958} response to the pandemic is to mitigate the mortality or infection risks. In this context, the majority of response design decisions are based on sequentially or hierarchically structured “if-then” or “cost-benefit” analysis drawing inferences from “cause-effect” scenarios. One way to reduce errors associated with the probabilities in calculations of efficiency or pay-offs from response strategies\cite{Shea2020, Bjornstad2020a,Drake2019}  is by supplanting it with scientific evidence through accumulating decision-support information of causal inferences from epidemiological data analysis and models. \\
\begin{figure}[htbp]
	\centering
	\includegraphics[scale=0.3]{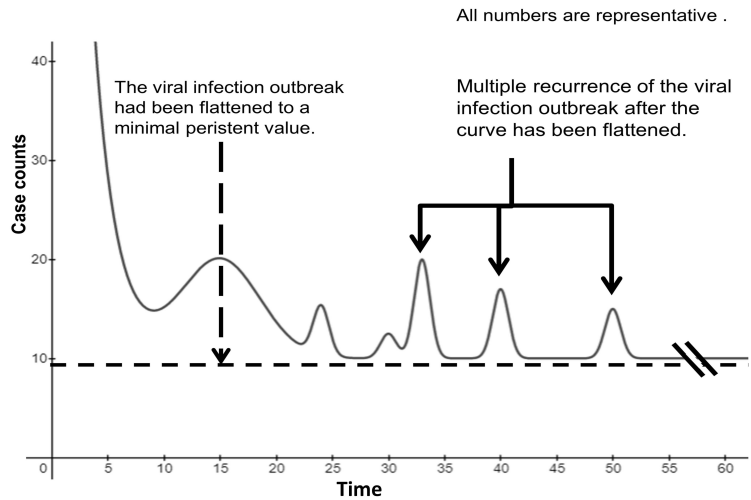}
	\caption{A hypothetical but probable scenario in which multiple recurrences in the incidence data of the country occurs due to state-levels dynamics.}
	\label{m1}
\end{figure}

{\noindent}Modeling is a process of inclusion and exclusion of parameters measuring real observables within a given context and there are many tacit assumptions, contexts, and trade-offs involved in such a model \cite{Iwasaki1994}. When using the model inferences to design any mitigation strategies, the feedback from convolution between contexts and parameters must be carefully analyzed to prevent spurious correlation or false causations entering the subsequent strategy designs. Modeling the ongoing epidemic/pandemic as a dynamical system is a drastically simple but useful level of abstraction which allowed us to quantify the dynamics of the trajectories of this system from multiple perspectives. The well-established epidemiological models based on compartmental subdivisions of population, in both stochastic and deterministic form, and taking into account the demographic diversity have been done rigorously. Emerging attributes like recurrences \cite{Whitman2020, Cacciapaglia2020, Hitz2020, Mollison1986, bartlett1956, Anderson1984,Schwartz1983,Gomes2014,Krylova2013, Korevaar2020},  synchrony in infection occurrences \cite{Cui2010, Cummings2004, Grenfell2001, Bjornstad1999, He2003}, super spreading \cite{Li2020,Wang2020}, technology-based contact tracing, etc. also have been studied and contextualized. However, still we are facing problems to control this ongoing pandemic.  \\

\noindent In this paper, we have modeled statistically and theoretically the development and evolution of the COVID-19 scenario in some of the individual states of India. One of the central motivations for the mode of analysis taken in the paper is the fact that multiple agents are influencing the pandemic dynamics as inferred from the incidence data. Figure \ref{m1} presents one of such scenarios in which there are small recurrences of the outbreak after the pandemic has been controlled to a certain minimal level. These recurrences may correspond to multiple wave like features which are seen in ongoing Covid-19 pandemic \cite{Hale}. One inference which could directly obtained is that the dynamics of the pandemic is necessarily stochastic, where, fluctuations in the dynamics may come from different factors including problems in the control policies, random interaction of individuals, environmental fluctuations, random climate change, change in the food habits etc \cite{Bittihn,Gardner} which may trigger the dynamics far from equilibrium \cite{Chanu} and may drive the system to various recurrences. Hence, fluctuation is one of the contextual factors to be taken into consideration when modeling the data of the pandemic incidence. Another example of causative inference could be that the manual interventions like lockdowns and time-dependent behavior of individuals’ mobility could also act as contextual factors. Thus, the central theme is set to analyze these causations or correlations between contextual factors and the model variables obtained from pandemic data.\\

\noindent The paper is organized to describe inferences from two conceptual data-based modes of analysis. The repertoire of statistical tools and methods, to draw inferences about the pandemic dynamics from data is huge and each tool is optimized for a certain class of constraints. The choice of the tools presented here is supplanted by their power to extract information about the latent factors which can then be interpreted safely with proper tests and checks.  Firstly, parametrized inferences and their respective critiques are given based on fitting the data to growth functions like the Gompertz’s, Richard’s function, and renormalization group theory functions. Secondly, analysis of the spatiotemporal data using Empirical Mode Decomposition (EMD) algorithm is performed to untangle the qualitative-quantitative effects of various latent factors on the dynamics of the pandemic. The data we have used might contain errors or biases that cannot be removed therefore the results are completely dynamic and may change with scenarios.\\
\begin{figure}
	\centering
	\includegraphics[scale=0.35]{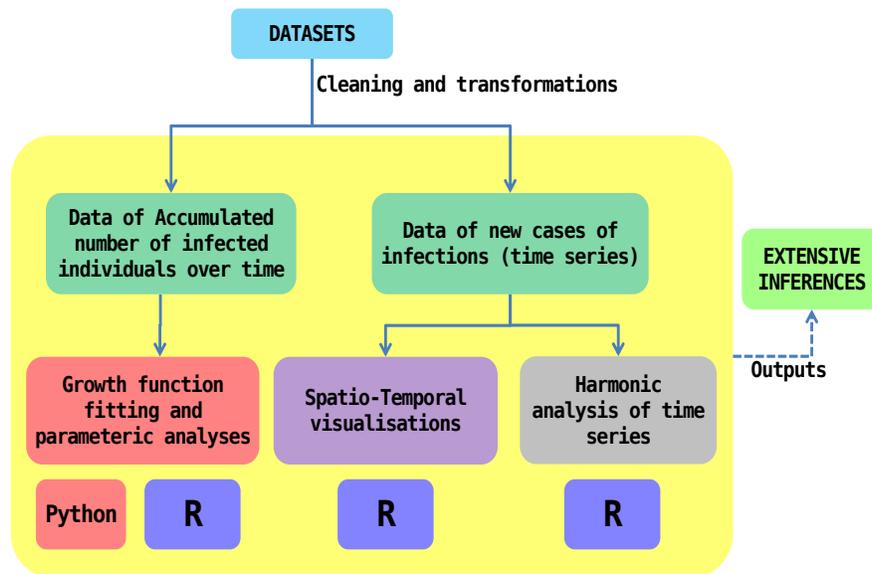}
	\caption{FLow chart used for data analysis}
	\label{m2}
\end{figure}

\section{Methods}
{\noindent}We describe the Covid-19 data acquisition from various data repositories, then to clean the data for analysis. We then explain various fitting procedures to the acquired data for possible analysis and prediction. Then we explain the method of harmonic analysis we used in this work.

\subsection{Data acquisition and cleaning} 
{\noindent}The epidemic/pandemic growth can be transformed to obtain inferences through several methods like fitting, visualizations, filtering etc of the data \cite{Lega}. The flowchart in Fig. \ref{m2} shows the approach we used in this work. The cleaning and transforming processes of the retrieved data are done to minimize the errors in subsequent operations of the methods on the data. The two main data sets used are: (a) the accumulated number of infected individuals denoted as $I(t)$, and (b) the new number of infected individuals denoted as '$n(t)$'. The data used in the work has been obtained from the dedicated Covid-19 database \cite{PRS} and then cross verified with the data present in the other government provided websites. The data used may contain many spurious human input errors thus inferences from the causative statements should be used taking these factors into account.\\

\subsection{Growth curves for fitting procedure}
{\noindent}Fitting of the growth functions to the accumulated number of infected individuals data $I(t)$   \cite{RICHARDS1959,Winsor1932,Bertalanffy1957,Tjorve2017,Somers1988} and then tracking the fit statistics in time may reveal growth-phase transitions of the pandemic which is both important in designing preemptive response programs and will provide a correlative-measure of the qualitative and quantitative effects of decisions taken to control the epidemic/pandemic. In this work, we analysed the the growth of the pandemic in Indian states by modelling the data through the fitting of the following growth functions:\\ 

\noindent \textbf{Gompertz function}\\
This function has a rich history \cite{Winsor1932,Tjorve2017,Consolini2020} of being used to fit multi-phase growth processes’ data. It is a three-parameter function given by,
\begin{eqnarray}
\label{Gom}
I(t) = a\mathrm{e}^{-\mathrm{e}^{b-{\gamma}t}}
\end{eqnarray}
where, $a$, $b$ and $\gamma$ are the three parameters in the function \eqref{Gom}. To analyse the nature of the data and growth of the pandemic, this function is fitted to the curve of the accumulated number of infected individuals with time. The three parameters control the following features of the curve: '$a$' controls the value of saturation plateau, '$b$' controls the inflection point and '$\gamma$' controls deviations from the logistic growth curves. The inflection point $T_i$ (stationary point) can be calculated by taking the derivative of the function \eqref{Gom} and equating it to zero, which is given by, $\displaystyle T_i=\frac{b}{\gamma}$. This inflection point in the fit parameters correlates with the turning point of the growth curve indicating the plateau of the growth curve is reached. If there are multiple inflections corresponding to multiple pandemic waves at $t_1$, $t_2$,...,$t_m$ ($m$ multiple waves), then $\displaystyle T_i(t)=\frac{b(t)}{\gamma(t)}$, where, $t_1\le t\le t_m$.\\

\noindent \textbf{Epidemic-Renormalization Group (e-RG ) theory growth function}\\
Renormalization group theory, which is used in various branches of  Physics and Mathematics \cite{Fisher,Jona}, can be applied to epidemiology \cite{Cacciapaglia2020a,Morte2020} to model and analyse the growth of the epidemic. The growth function given by this theory is given by,
\begin{eqnarray}
\label{rg}
ln[I(t)]=\dfrac{a\mathrm{e}^{{\gamma}t}}{\mathrm{e}^{{\gamma}t}+b}
\end{eqnarray}
where, $a$, $b$ and $\gamma$ are the three parameters involved in this growth function. In the same way as is done in Gompertz function case, the one wave inflection point is given by, $\displaystyle T_i=\frac{1}{\gamma}   ln\left[ \frac{b}{2} \left( a+\sqrt{4+a^2}\right) \right]$. For $m$-multiple wave scenario, the multiple inflection points are given by, $\displaystyle T_i(t)=\frac{1}{\gamma(t)}   ln\left[ \frac{b(t)}{2} \left(a(t)+\sqrt{4+a(t)^2}\right) \right]$, where, $t_1\le t\le t_m$.\\

\noindent \textbf{Richard growth function:}\\
Richard’s function \cite{RICHARDS1959,Somers1988} is a four-parameter growth function which can be used to study the characteristics and growth properties of an epidemic/pandemic by fitting the epidemic/pandemic data. The function is given by,
\begin{eqnarray}
\label{rf}
I(t)= \frac{a}{\left[1+ {{e}^{-b(t-d)} }  \right] ^{\frac{1}{\gamma}} }
\end{eqnarray}
where, the growth parameters are $a$, $b$, $d$ and $\gamma$. The main underlying assumption of this function is that there is a single maxima peak in the daily new cases data i.e. $n(t)$. For single wave epidemic/pandemic, the inflection point is given by, $\displaystyle T_i= d-\frac{ln \gamma}{b}$. However, for $m-$multiple epidemic waves, there will be $m$ such inflection points given by, $T_i(t)= d(t)-\frac{ln \gamma(t)}{b(t)}$ for $t_1\le t\le t_m$.\\

\subsection{Harmonic analysis of spatiotemporal data}
\noindent The theory of drawing causal inferences from a time series data through spectral methods is a powerful methodology of extracting patterns and characteristics of the data based on the decomposition of the signal into inherent modes \cite{Huang1998}. In the context of a pandemic, the data of daily new cases $n(t)$ (i.e. a time-series data) can be analyzed using harmonic analysis to understand the nature of the epidemic growth patterns \cite{Anderson1984,Huang1998}. \\

\noindent Geo-encoded temporal data of incidence \cite{Bjornstad1999, Bjornstad2001,He2003,Cacciapaglia2020} gives a powerful way to visualize the physical spread of the pandemic through the states. Coupled with the information provided from statistical analysis of the spatiotemporal (time-series) data, the harmonic analysis can provide both correlative and causative insights into the relational mappings of the contextual factors and the data-derived inferences. Correlations among this geo-coded spatiotemporal data can be used to visualize the measurable population attributes like synchrony in spreading of the disease with time and spatial regions \cite{Cummings2004}. These features have been observed in various diseases spreadings  \cite{Hitz2020,Grenfell2001} and their quantification could help in risk mitigation and design of response by the appropriate administrative authorities.\\

\noindent Empirical mode decomposition(EMD) \cite{Huang1998,Stallone2020} is a robust method by which we decompose the given time series into inherent signals with different frequencies known as intrinsic mode functions (IMF). The method is done by a process of sifting (Please refer to the Supplementary Information section). The process can be thought of as analogous to a filtering technique where multiple frequency components in a time-series can be extracted step-by-step in a data-adaptive manner. One of the major advantages of this method is the fact that all analyses are done a posteriori and based on data solely.

\section{Results}

{\noindent}The retrieved data from the Covid-19 dedicated database in Indian context \cite{PRS} is cleaned and applied the methods mentioned in the \textit{Methods} to analyse the data. The results are discussed as in the following.

\subsubsection{Mapping of growth phases of Covid-19 on a country-wide scale }

\begin{figure*}[htbp]
	\centering
	\includegraphics[scale=0.125]{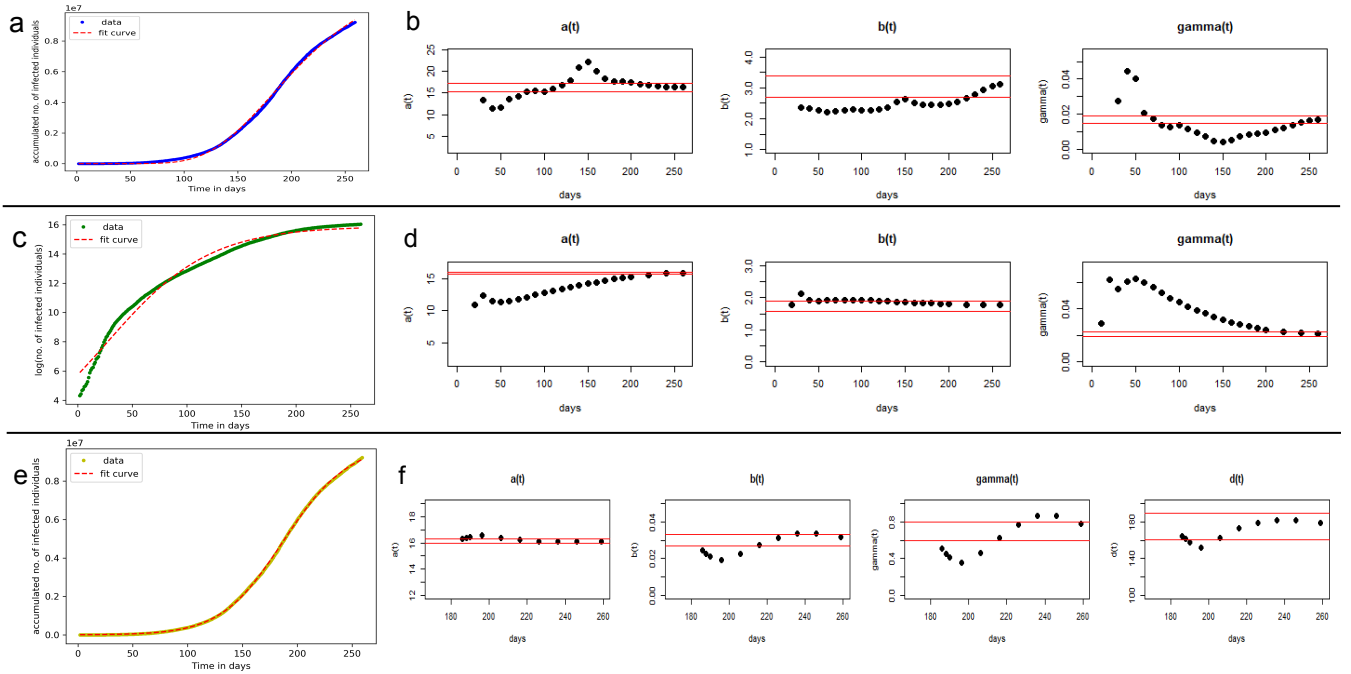}
	\caption{Fitting of countrywide accumulated no. of infected individuals data ( \textbf{I(t)} ) to growth functions and tracking the temporal changes in fit parameters to measure the reaction of the disease spreading to interventions. (\textbf{a-b})Gompertz growth function and temporal changes in fit parameters, (\textbf{c-d}) epidemic Renormalisation Group theory growth function and temporal changes in fit parameters, (\textbf{e-f}) Richards growth function and temporal changes in fit parameters.}
	\label{m3}
\end{figure*}

\noindent The data of the cumulative number of infected individuals I(t) for the whole country for 260 days from the period of March 2020 to November 2020 was fitted to three growth curves as shown in the Figure \ref{m3}. The fitting of the three growth curves mentioned in \textit{Methods} has been done using Python code\cite{py}. All the fitted curves on the data as shown in the Figure \ref{m3} provide a picture of a smooth growth with a saturation reached after some time. This fitting results provide a good basis for comparative modeling of the pandemic with compartmental models.\\

\noindent The fit parameters are in the table provided in the supplementary information.  The first panel on top of the Figure \ref{m3} shows the data fit with a Gompertz growth function. The fit parameters show a time-dependent behavior as shown in the three panels in the row of Figure \ref{m3}, (b). The parameter $ a(t) $ encoding the number of infected individuals increases rapidly but as strict lockdowns are enforced, it peaks and then saturates to a lower value correlating with the fact that the intervention measures were working to contain the spread of the disease. A major transition in the growth phase of the pandemic is prominently shown by the existence of inflection points in the plots of the time-dependent parameters. This feature is seen in the behavior of all the growth models. The correlation between the parameters also changes its sign after the inflection point has occurred e.g. the correlation between $ b(t) $ and $\gamma(t)$ goes from negative to positive after an inflection point occurring around 150-160th day. If one had analyzed the data from the start of the pandemic till the 110th day, one could easily notice that it could predict the time when there would be a peak in the number of cases with an error of $\pm$ 7 days.\\

\noindent In case of fitting with the RG function as shown in the second panel of Figure \ref{m3} (c-d) the parameter a(t) shows an increase correlating with the facts that increase in the number of next-neighbor infections resulting from the increase in diffusion mobility amongst the population due to uplifting of strict lockdowns on 80th day. The parameters b(t) encodes information about the offsetting time in the occurrence of the turning point of the growth curve. The parameter $\gamma(t)$ encodes information about the flattening of the growth curve. Thus lowering of $\gamma(t)$ indicates the lockdown measures were effective in the control of the pandemic spread. The parameters a(t) and $\gamma(t)$ are correlated. Again a similar change in behavior of the correlation between parameters is seen after the occurrence of an inflection point.\\

\noindent The bottom panel of the Figure \ref{m3} (e-f) shows the fitting of data with the Richard growth function. The main difference between this model of fitting from the others is that the data has to be fitted from the 186th day after the start of the pandemic. The fit could not be performed piecewise as the others. Though this lacuna was present, the fit parameters were accurate(within a certain error limit) in obtaining the information about when the peak of the daily number of new cases occurred.\\

\noindent The growth curves of Gompertz and RG functions were fitted for the data of states of Delhi, Maharashtra, and Karnataka and have been shown in the supplementary information[Figures 5-7]. The fit statistics show similar behavior as seen for the country-wide data. This feature of behaviors of fit statistics supplants the claim that the dynamics of the disease on a country-wide scale emerges from local state-wide dynamics. Comparatively, one can also see which state is far better in the case of control strategies and response designs. These can serve as crucial precedents in states with the pandemic in an early stage of growth.\\

\subsubsection{Spatiotemporal Mapping of growth of Covid-19 in individual states}

{\noindent}The systematic analysis of the data with the visualization procedure can be interpreted to provide insights of the nature of the spread of the pandemic and correlational information about the effects of the interventions and response strategies. The results are presented here from the analyses mentioned in the previous section i.e. firstly the fit statistics from the growth function fitting and then the harmonic analyses of the spatiotemporal data. Some of the inferences are: \\
\begin{figure*}
	\centering
	\includegraphics[scale=0.65]{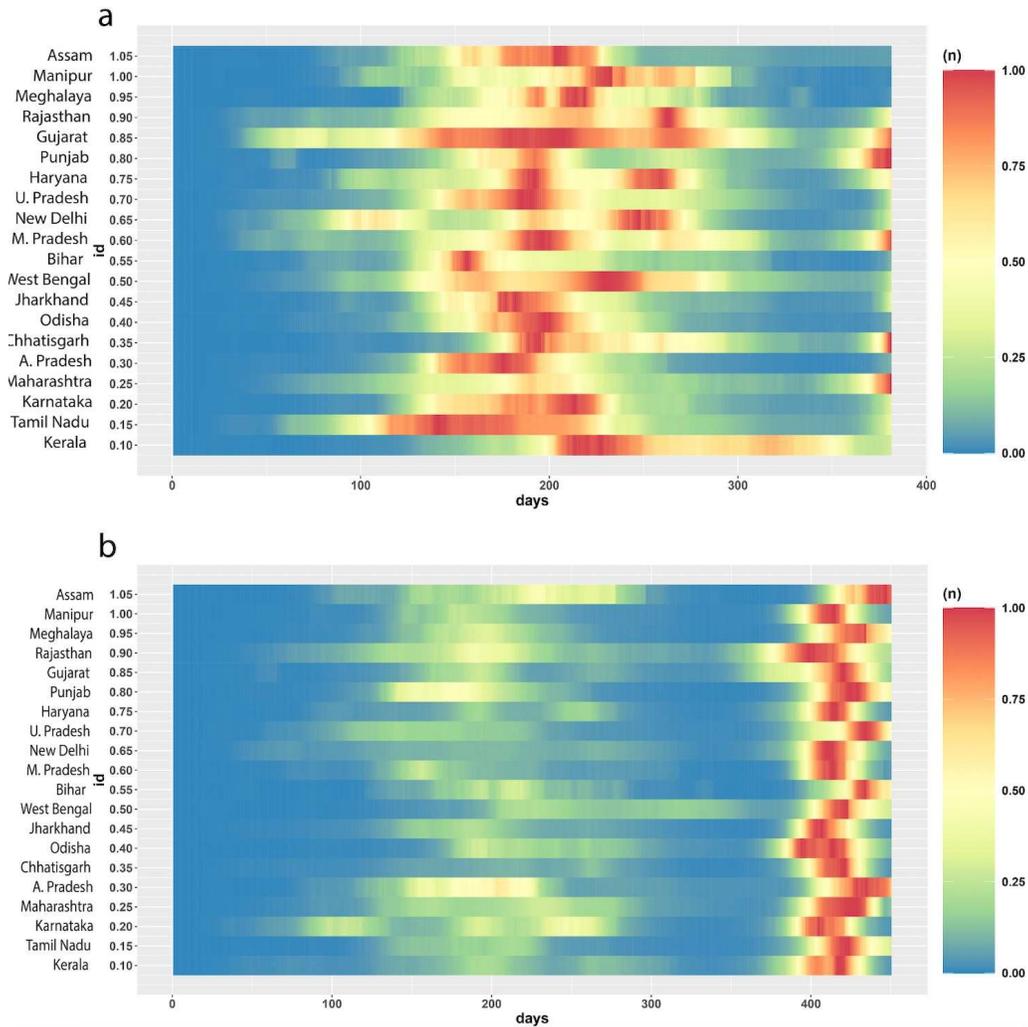}
	\caption{ Sequential analysis of Spatio-temporal plots showing the evolving wave like features in the dynamics of the pandemic. (a)Plot showing the normalized data of states 381 days after the first infection incident. (b)Plot showing the normalized data of states 450 days from the first infection incident.}
	\label{m4}
\end{figure*}
\noindent 1. The wave speed in essence depends on the various interactions and demography effects and thus changes the speed (accelerates or deaccelarates) of the advance of the disease in the population habitat\cite{Birzu2018} . Since the above model is also analogous to the modelling of invasion of infection into a susceptible population the rate of invasion can be seen depending on the perturbation term ie.e the interaction terms arising due to various effects like inherent variation of the degree of susceptibility of the population to the viral infection.

\noindent 2. Diffusion plays an important role by creating instabilities hence it raises the need to monitor the mobility of infected popualtions in real time as these will be sequentially used as initial conditions for model simulations to infer parameters.

\noindent 3. We reiterate that the model presented is an highly abstracted and simplified form of a large numer of complex factors, trade-offs such as demographic variation of susceptitbility to infection, time-delays caused due to discrepancies between onset of infection, testing of the individual and finally reporting, population heterogenity and density etc. (all these factors can play a confounding role while inferring parameters). 

\noindent 4 Inspite of the simplifications as seen from the wave picture, an essential inference can be drawn is that to stop or slow the advance of the disease we need to create as given by Fisher's \cite{FISHER1937} terms "centres of extinctions" and identify then isolate the "centres of multiplications" by building buffer zones around them(to reduce the motility ). \\

\noindent Latent features in the time-series data can be extracted using harmonic analysis. Without taking into account the spatial regions this temporal information of the dynamics of the pandemic is only half the picture. The analysis and subsequent visualization of the geo-encoded data are analogous to the contextualization of the information provided by the harmonic analysis to the concepts of demography, population scales, etc.\\

\subsubsection{Geo-encoded harmonic analysis shows synchronous incidence patterns}

\noindent The time-series data as shown in the Figure \ref{m4}(a) have been geo-coded and plotted as a Hovmoller type plot in the Figure \ref{m4} (b) .  As can be seen in Figure \ref{m4} (b) there is a signature of a traveling wave-like feature present across the country. To further cross-analyze whether this spatio-temporal dynamics is an artifact of visualizations, cross-correlation (CCF) analysis is performed using R software\cite{R}.
In the Figure \ref{m4} (b), the states are arranged spanning from north to south.  As can be seen in each panel the peaks of the incidence occur in the same months. This feature can be labeled as countrywide incidence synchrony. One of the causes of this synchrony can be the simultaneous uplifting of lockdown across the country and a correlated increase in the diffusive-mobility of individuals. Further, as the pandemic dynamics in states are visualized by clubbing them together local next-neighbor state-wide synchrony is also prominent.\\

\subsubsection{EMD and CCF analysis}

\begin{figure*}
	\includegraphics[scale=0.1]{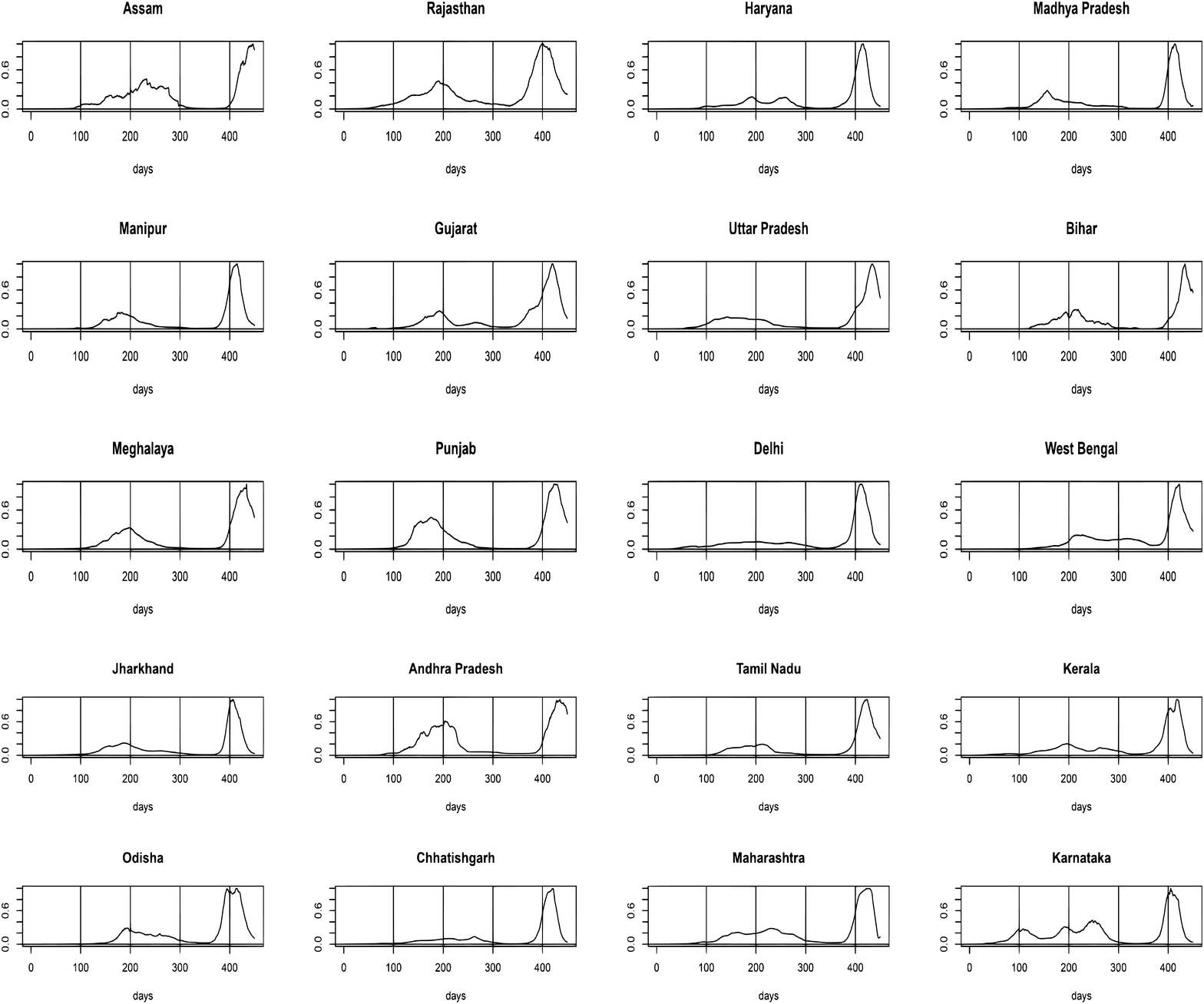}
	\caption{Plots of the time series data of various states after EMD analysis.}
	\label{m5}
\end{figure*}

\noindent To verify whether these features are due to the architecture of visualizations only, EMD analysis was performed on the time-series data which yielded Intrinsic Mode Frequencies (IMFs) present in the data. These IMFs are then summed up and plotted in Figure \ref{m5}. A prominent and similar feature in the plots is the existence of multiple maxima or an initial decrease in the function .The behavior depicted in these plots again can be correlated to the increase in the mobility amongst the population after simultaneous uplifting of lockdown measures (on the 80th day) and the variant evolution.

\begin{figure*}
	\centering
	\includegraphics[scale=0.8]{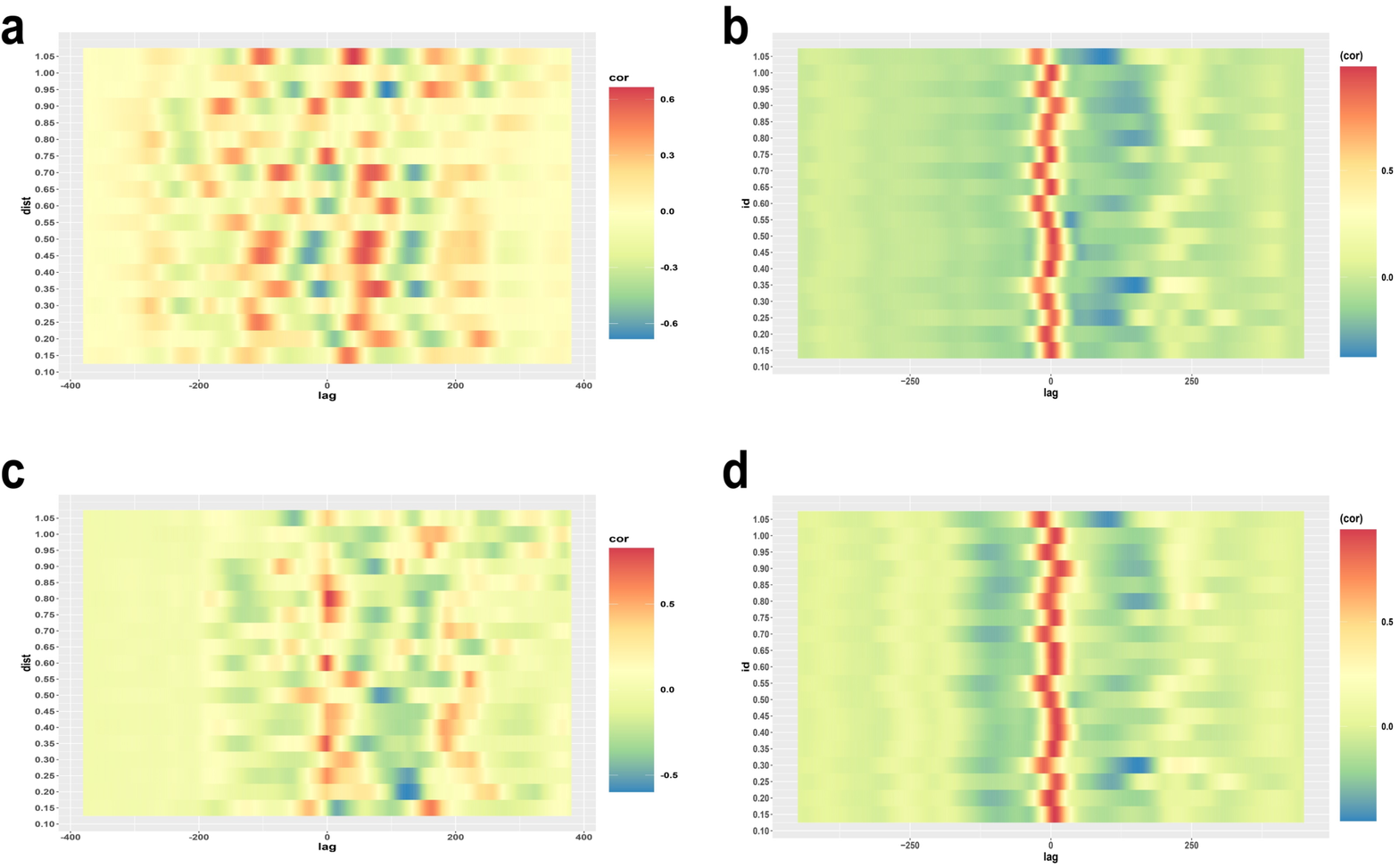}
	\caption{(a-b)CCf of time series data of various states with Delhi time series data.(a)Plot showing the correlation values for 381 days after the first infection incident. (b) Plot showing the correlation values 450 days from the first infection incident. (c-d) CCf of time series data of various states with Maharashtra time series data.(c)Plot showing the correlation values for 381 days after the first infection incident. (d) Plot showing the correlation values 450 days from the first infection incident.  }
	\label{m6}
\end{figure*}

\begin{figure*}
	\includegraphics[scale=0.1]{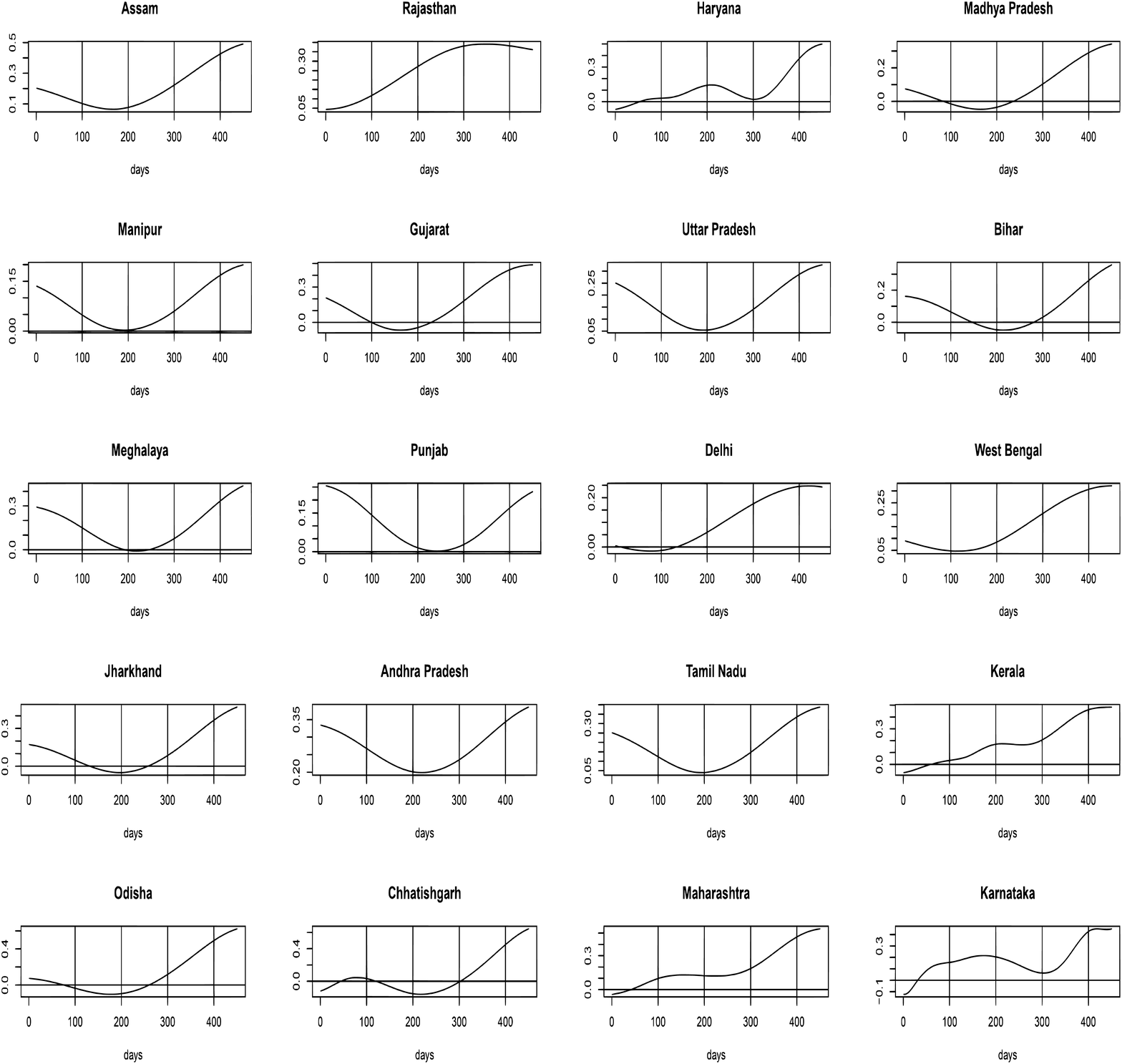}
	\caption{Plots of the residuals after EMD analysis of the time series data of various states. These represent latent trends present in the data which are not observed when simply plotted.}
	\label{m5}
\end{figure*}

\noindent At this junction, one hypothesis is introduced to simplify the complexity associated with the causes of this wave-like feature. Since all states have open borders,  a high variation in population density, and a diverse set of demographic variables, they can be described as coupled components in a complex system. In this context, one can hypothesize the presence of varied sources of traveling waves. The patterns observed in the cross-correlation between the data of Delhi, Maharashtra, and the other states as shown in the Figure \ref{m6} (a-d) .The spatial extent of this wave is very large and can be said to cover almost the whole country as seen in the plots  Figure \ref{m6} (a-b) and the Figure \ref{m6} (c-d) by the presence of a strong correlation across time between the time series data of Delhi, Maharashtra with rest of the states. 
\begin{table*}
	\begin{threeparttable}
		\caption{Summary table of Inferences.}
		\centering 
		\begin{tabular}{ p{3cm} p{1cm} p{13cm}  }
			\hline 
			\hline			
			Mode of analysis&  & Inferences  \\	
			\hline		
			\hline
			\textbf{Traveling wave} & & Identification of patches of high infection rates based on speed of advance of the disease wave and then creation of buffering zones to reduce the mobility or vaccination drives in or near by areas where past data shows high degree of infections.\\
			\hline
			&&\\
			\textbf{Growth functions fitting} & & 1. If the data is fitted and the fit statistics are tracked piecewise (on a weekly or daily basis), one could infer when the peak of the daily number of new cases has occurred. This is critical for the agencies involved as it may allow for adaptive changes in policies and interventions taken e.g. to design and decide for strict or moderate lockdowns, the degree upto which the bus/train/airport terminals etc. are opened or closed.  \\
			& &\\
			& &	2. Each growth function has it's own advantage. The information provided could be compared amongst models or separately stored as a repertoire in a data-base. This pluralistic mode of analysis may be used to make components of a Decision Support System(DSS) during designing of sequential response strategies. \\
			&&\\
			&& 3. The behavior of the fit statistics are similar when done on state wide scales and on country wide scale.\\ 
			&&\\
			\hline
			&&\\
			\textbf{Harmonic analysis}  & & 1. Spatio-temporal synchrony is an important driver of infection incidence across country. To reduce the speed of the invasion of the disease outbreak measures could be taken to break down this synchrony   \\
			& &	2. Dense transportation hubs such as port cities or densely populated cities should be identified and monitored as they could act as centres for emanating travelling waves. \\
			\hline
			
		\end{tabular}
		
	\end{threeparttable}
\end{table*}

\section{DISCUSSION}
{\noindent}The effects of control strategies of Covid-19 pandemic are reflected in the observed data and the analysis of these data could reveal the magnitude of the importance of the strategies and other properties. Implementable response strategies take their inputs from precedents (if present) and inferences from iteratively validated models. Models, both qualitative and quantitative are built on tradeoffs. Thus inferences from models are prone to biases that become significant at later stages of a sequential response program. Similarly, precedents contain entangled information about context-based solutions \cite{Wauchope2020} that can create long term problems if the information is used without proper analyses of cause and effects. We have tried to factor these in by correctly stating whether the inference is a causative one or a correlative one. The dynamic character of the parameters as elucidated from the above analyses may provide probabilistic information about the effects of the interventions taken \cite{Lilleri2020}. Time to time analysis of the growth of the disease outbreak through the fitting of curves and tracking the fit parameters must be undertaken for a continuous adaptive mitigation program. \\

\noindent The wave-like features illustrate the nature of the invasion of the disease into suspected population pockets driven by an increase in diffusion mobility of individuals. Here a two-pronged methodology can be adopted. Non-linear dynamics of synchrony in population growth can be helpful to model the epidemic development in terms of number. There is a well-established literature on this subject which can help to model the effects of inherent demographic stochasticity in the growth of epidemic inside the country. These analyses can be performed on national or statewide scales. One future work is mapping the edge of these waves \cite{Panja2003} where the number of infected individuals is small and the character of the wave is dominated by a pulled front wave \cite{Birzu2018}. Modeling these aspects together with the population heterogeneity and influx can help understand the dynamics of the epidemic development. \\

\noindent Another factor that adds to the stochasticity in the dynamics is the evolution of the virus. Thus, genetic changes in the virus and its correlation with the infected population heterogeneity must be tracked in a Spatio-temporal manner \cite{Anderson1984, Langwig2017}. The inferences from the growth curve fitting, harmonic analyses together with inferences from genetic analyses could provide a holistic picture of the growth and evolution of the pandemic.

\end{document}